# Graphene field-effect transistors for sensing ion-channel coupled receptors: towards biohybrid nanoelectronics for chemical detection

*Océane Terral, Guillaume Audic, Arnaud Claudel, Justine Magnat, Aurélie Dupont, Christophe J. Moreau, Cécile Delacour\**

Océane Terral, Arnaud Claudel, Cécile Delacour
Institut Néel, University Grenoble Alpes, CNRS, Grenoble INP, 38000 Grenoble, France
E-mail: cecile.delacour@neel.cnrs.fr

Aurélie Dupont
LIPhy, University Grenoble Alpes, CNRS, 38000 Grenoble, France

Guillaume Audic, Justine Magnat, Christophe J. Moreau
Univ. Grenoble Alpes, CNRS, CEA, IBS, F-38000 Grenoble, France

## ABSTRACT

Graphene field effect transistors (G-FETs) have appeared as suitable candidates for sensing charges and have thus attracted large interest for ion and chemical detections. In particular, their high sensitivity, chemical robustness, transparency and bendability offer a unique combination for interfacing living and soft matters. Here we have demonstrated their ability to sense targeted biomolecules, by combining them with ion channels-coupled receptors (ICCRs). These receptors have been naturally or artificially expressed within living cell membranes to generate ion fluxes in presence of chemicals of interest. Here, we have successfully combined those biosensors with G-FET array which converts the bio-activation of the ICCRs into readable electronic signals. This hybrid bioelectronic device leverages the advantages of the biological receptor and the graphene field effect transistor enabling the selective detection of biomolecules, which is a current shortcoming of electronic sensors. Additionally, the G-FET allows to discriminate the polarity of the ion fluxes which otherwise remains hidden from conventional electrophysiological recordings. The multisite recording ability offered by the G-FET array rises numerous possibilities for multiscale sensing and high throughput screening of cellular solutions or analytes, which is of both fundamental and applied interests in health and environment monitoring.

## INTRODUCTION

There are several approaches to measure the content of analytes (e.g. spectroscopy, chromatography, mass spectrometry, electrochemical methods, immunoassays, molecular biology techniques). The choice of method depends on factors such as the nature of the analyte, the required sensitivity, the sample matrix, the available instrumentation and the operation conditions. Among them, potentiometric electronic devices such as microelectrodes or field effect transistors enable long lasting, real time and on-site monitoring. They are compact, easily integrated in portable or wireless systems, compatible with microfluidic and large-scale implementation. More recently, successful demonstrations have combined them with promising biological sensors, comprising cells, receptors, peptides, nanovesicles that have been recently review.[1] Because of the broad range of ligand-gated proteins, their integration on electronic sensors is of particular interest to identify and enlarge the number of detectable chemicals. Specially, ion channels (ICs) generate fast and large flow of ions (µs to ms, and up to millions per second respectively) that might be sensed by ion-sensitive field effect transistors (ISFETs) and converted into readable electrical signals. However, the number of endogenous ligands recognized by vertebrate ligand-gated ion channels (LGICs) is very small (< 10),[2] which limits their applications in the field of biosensing.

To expand the repertoire of recognized ligands, two approaches have been explored in the present study with non-vertebrate ICs and artificial ICs with exogenous receptor moieties. In particular, a highly conserved odorant receptor co-receptor Orco from *Drosophila melanogaster* (fruit fly, a common scientific model), generates high current amplitudes in response to few synthetic ligands in Xenopus oocytes,[3] and it was consequently used as a first model in this study. On the other hand, artificial ICs have been created by fusing ICs with G protein-coupled receptors (GPCRs),[4,5] which are called ion channel-coupled receptors (ICCRs). Integrated in cell membranes, ICCRs are suitable for chemical sensing operation in a physiological fluid, taking advantage of specific combinations of ligand-receptor for real-time detection and identification of various biomarkers or chemical compounds in general. When the molecule to be-sensed (ligand) binds to the protein receptor, conformational changes of the protein trigger the opening or closing[6] of ICs that are coupled, as illustrated within figure 1. Because of high gradients of ion concentration between intra and extracellular media, this modulates ions fluxes through the lipid membrane that can be detected by electrophysiological methods.

The initial demonstration of ICCR biosensing[2] was carried out on Xenopus oocytes which are a standard heterologous expression system for ICs. These living cells have several advantages for operation conditions in standard laboratories. Their optimal temperature is close to room temperature (19°C) (no temperature control equipment), no $CO_2$ is required, they survive for days in minimum salts buffers and up to hours in a peeled sate, and they are quiescent (no cell division or change of morphology during the experiments). The detection of ligand-induced currents is usually performed in whole cell configuration with the two-electrode voltage-clamp (TEVC) method.[7] While TEVC robots have been developed in 96-well plate format since, the level of throughput could be largely increased with multiplex recordings, as well as avoiding invasive cell manipulation should significantly increase the observation time-window. To make progress towards this objective, extracellular measurements have been developed that do not alter the cell membrane or intracellular fluid and enable long lasting recording of cell activity during weeks. Array of electronic devices such as microelectrodes or microtransistors could be used, being compact, easily integrable in microfluidic circuit and wireless systems, and compatible with large scale implementation on any kind of substrates, for non-invasive and large-scale screening of the biosensors with user-friendly sensing platform.

The drawback is that extracellular variations of potential or ion concentration are lower than intracellular changes,[8] and thus more difficult to sense. As illustrated figure S1, this result from the high impedance of the cell membrane, the leakage currents to the grounded bath, or the charges screening in the solution. The typical ion concentration of the bath being about 100mM, the Debye length is significantly reduced (1nm, figure S2) compared to ultra-pure water. Moreover, the ICs of the oocytes are embedded in a membrane which is rather folded at the microscale such that ICs could be few microns away from the device and at least above the Debye screening length (1nm at 100mM ionic strength of the electrolyte solution, figure S2). In addition, the number of detectable ICs is significantly reduced in comparison with whole cell TEVC measurements, reducing the detection efficiency. These limitations might be overcome by overexpressing locally the ICs of interest to increase their density around the recording site. Also, the use of highly sensitive and biocompatible materials such as graphene is a way to promote both ISFET performance and the adhesion of the cell membrane to ensure high sealing resistance.

Due to their high sensitivity, affinity with the cell membrane and chemical inertness, graphene field-effect transistors (G-FETs) have appeared as a promising candidate for such ion detection in physiological fluids. This 2D honeycomb lattice of carbon atoms exhibits high charge-carrier mobility and promotes the adhesion of cell, providing a high sensitivity to extracellular potential changes and to reach high-frequency operation regime. This is further supported by the chemical inertness of monolayer graphene, which enables to expose the transistor channel directly to liquids at the closest vicinity of the environment to be sensed, such an ultrathin electrical double layer stands alone at the interface with the liquid or the cell to-be-sensed. This unique property of combining high mobility of charge carriers and an atomically thin interfacial layer, enables to reach higher sensitivity and frequency regimes than its counterpart silicon or carbon nanotube FETs. First demonstrations were achieved for sensing ion channels,[9] bioelectric signals[10] and ions in solution.[11] Since, the use of G-FETs have been extended for sensing analyte composition,[12,13] DNA,[14] pH,[15,16] virus[17] and ionic signals generated by electrogenic cells such as neurons[18,19] or cardiomyocytes,[10,20]. These examples illustrate the broad range of possible applications for graphene-based ISFET in the fields of health and environment. Enhanced performances such as high selectivity and sensitivity could be achieved by functionalizing graphene directly with bio-recognition elements, such as enzymes or antibodies[21] or ionophore membrane[12]. However, this limits the detection to the molecules where their bio-recognition elements exist and can be used to functionalize graphene. Also, this restricts the detection of a single and predefined species per GFET. Recent attempts such as the introduction of the Hall or field effects and the use of pristine graphene[22,23] might overcome this limit.

Here, bare graphene FET were directly exposed to programmable ICCR bringing promising outcomes (1) for real-time and extracellular sensing of ion channels at the microscale within living matter, and (2) for enlarging the available receptors among the more than 800 human members. The two being of fundamental and applied interests. The field effect detection of ICCRs by G-FET is described in figures 1a and 1b. The ions adsorbed on the graphene induce an electrostatic gating. The surface potential is changed which tunes the Fermi level. As a result, the charge neutrality point $V_D$ is shifted by a value, which depends on the ion concentration. Therefore, the ionic concentration of a solution can be deduced by monitoring the gate potential $V_{LG}$ needed to reach the charge neutrality point. This technique is well adapted to detect the proton concentration of an electrolyte and to determine pH of solutions, as reported in several seminal works.[16,24,25]

Considering a perfect single crystal of graphene, its small amount of dangling bonds should not favour ion adsorption, which is a primary key for ion sensing. However, it was shown few years ago that intrinsic defects of the CVD-grown graphene-monolayer (e.g. sp3-hybridized carbon atoms) constitute sites of high chemical reactivity for enhanced sensitivity.[26,27] Those specific sites favour a charge transfer between the solution and the graphene which allows to reach high sensitivities to detect adsorbed ions or molecules in liquids[25,26] and with biological membranes.[23]

Here, we show the application of graphene microtransistors for sensing ligand-activated IC activity within Xenopus oocytes, and their potential for implementing future generation of hybrid bioelectronic sensors.

## RESULTS

**Manufacturing ion-sensitive graphene field effect transistors.** Because of its chemical stability graphene monolayer can be directly exposed to the cell and liquid media increasing considerably the devices performance in comparison with more conventional silicon or indium tin oxide (ITO) materials. Indeed, the FET sensitivity is inversely proportional to the gate thickness, which is about 1 nm and 10 nm for graphene and silicon FET respectively. Additionally, graphene nanomaterial can be reported on any kind of substrate including transparent ones, which provides optical access to the biological object of interest by using standard transmission microscope. Here, the monolayer graphene was obtained by chemical vapor deposition (CVD) and transferred from the copper growth substrate onto sapphire by wet transfer technique. Then arrays of graphene field-effect transistors have been fabricated in a clean room microfabrication facility (details in materials and methods).

Figure 2a shows typical sample and zooms on a G-FET to illustrate the high quality of the resulting graphene channels directly exposed to the analytes. Only few residuals of the fabrication process are observed, and nanoscale wrinkles can already be observed as highlighted by the arrows. The atomic force micrograph confirms the overall quality of the graphene transistor channel (figure 2b). The monolayer is continuous with typical wrinkles of about few nm height (section profile in blue) and a mean roughness of about 458 pm which reveals the high quality of the monolayer graphene after all fabrication process. This value is below the expected Debye length (1nm at 100 mM, figure S2) which demonstrates the high quality of the fabricated G-FET and their potential for ion sensing.

Raman spectra have been acquired on the pristine graphene and on the final sample to assess its quality in term of contamination and defects (figure 2c). Single spectra were acquired on different regions of the graphene channel using a laser excitation of 532 nm (2.33eV) and low incident power of 1 mW. Each spectrum results from 10 accumulations of 5 second acquisition. Two main peaks are observed at $1591 \pm 4$ cm$^{-1}$ and $2693 \pm 2$ cm$^{-1}$ corresponding to the G and 2D bands related with graphene. The intensity ratio $I_G/I_{2D}$ =0.27, the G-band peak frequency and full width at half maximum (FWHM) of the 2D peaks ($\Delta\omega$=46 cm$^{-1}$) are as expected for monolayer graphene.[28,29] A single peak in the 2D-band stands for a monolayer graphene. Moreover, the D-band peaks (1348 cm$^{-1}$) reveals the presence of defects within the pristine graphene and the G-FET channel. Comparing it with the D' peak (1626 $\pm$ 3 cm$^{-1}$), which appears as a small shoulder of the G peak, enables to probe the nature of defect,[30] being mainly vacancy, impurity, and topological defects. In particular, the intensity ratio $I_D/I_{D'}$ =1.2 indicates a majority of sp2-hybridized boundary defects that can be found within grain boundaries for polycrystalline CVD-grown graphene. Those defects are expected to support the field effect detection for sensing ions.[23,25,26]

Field-effect measurements have been performed to assess the G-FET sensitivity to potential variations in liquids (figure 2d). The drain current $I_D$ was monitored while varying the liquid gate potential $V_{LG}$ from -600mV to 600mV with a quasi-reference platinum electrode immersed in phosphate-buffered saline solution (150mM PBS, pH=7.4). The $I_D$ versus liquid gate voltage $V_{LG}$ reveals the expected ambipolar field effect behaviour of G-FET, with a slight p-type behaviour at zero gate voltage, which was previously confirmed by Raman micro-spectroscopy for pristine CVD grown monolayer.[31] The derivative of the liquid gate transfer curve enables to extract the transconductance $g_m$ (red curve, figure 2d) which is directly related to the G-FET sensitivity S=$g_m$/V$_{DS}$ and enables to set the proper operation setpoint. Here, the maximum of sensitivity is around 1 mS/V (at $V_{LG}$~150mV).

**Ion sensing with the graphene FET.** Beyond their interest for the present study, which targets the detection of inward rectifier potassium channels (Kir), potassium channels are also expressed in many cells, which makes its detection relevant for other applications. To that end, the transfer curves $I_D$-$V_{LG}$ were measured in KCl electrolytes and for several ionic concentrations. The G-FET conductance was monitored while varying the gate potential $V_{LG}$ applied in the ionic solution whose volume and ion concentration were controlled. Starting from deionized water, small amount of highly concentrated ion solution was added in the electrolyte to gradually increase the ion concentration. Typically, we proceeded by successive injections of 1M KCl solution. Figure 3a shows the evolution of the transfer curve as function of several ionic concentrations, ranging from 0 to 300 mM (yellow to blue curves).

By increasing the KCl concentration, the charge neutrality point shifted towards negative values, indicating a negative doping of the graphene channel that could result from either chemical or electrostatic interactions. The adsorption or physisorption of K$^+$ at the vicinity of the G-FET channel or the increased ionic strength of the electrolyte[11,32] can both induced the accumulation of electrons within the transistor channel. The G-FET sensitivity for potassium detection is extracted from the dependence of the Dirac shift $\Delta V_D$ with the ionic concentration, and can reach -60 ± 3 mV/decade (figure 3d). This sensitivity with bare graphene is quite similar to what is obtained for membrane-coupled G-FETs[12]. This demonstrates the high performance of unfunctionalized graphene microtransistors which keep the ability to detect several ion species on a same device. To assess the dynamic of the G-FET response, the drain current was monitored while increasing KCl concentration in time (figure 3b and c). The G-FET was polarized at $V_{DS} = 80$mV and the liquid reference was fixed at $V_{LG}=400$mV or -500mV (resp. figure 3b and c) to optimize the current response of G-FETs to variations of concentration (see figure 3a). The initial solution was 200µL deionized water. Then successive amounts of KCl stock solution were added to increase the concentration of ionic charges in solution (figure 3b and c). At $V_{LG}=400$mV (resp. -500mV), the majority of charge carriers are electrons (resp. holes). Each injection of few µL of highly concentrated KCl solution induces a positive (resp. negative) change of the drain currents which amplitude increases with the amount of injected KCl ions. The dynamic response of G-FET to KCl injection is characterized by two phases: one rapid increase of the drain current until a maximum transient value and a second phase of slow decrease to reach a stabilized current value. The maximum value corresponds to the G-FET response to an instantaneous injection of the stock solution, inducing a highly concentrated ion flux. The stabilized value is related with a permanent regime corresponding to a homogeneous ion concentration in the electrolyte. Both values are not proportional to the injected ion quantity or related ionic concentration. Indeed, it seems that variations of current increase as a logarithm of the ion concentration (figure 3e, red circles). However, these current variations are comparable to what is expected by extracting drain current values from the transfer curves as function of the KCl concentration, and at fixed gate potential $V_{LG}$, (figure 3e, blue circles). This effect is directly related to the induced variations of potential at the graphene surface which one is not linear as a function of the surface charge density but follow the Grahame equation: $V_S = \frac{2k_B T}{e}\sinh^{-1}\left(\frac{en_c}{\sqrt{8\varepsilon_0 \varepsilon_r k_B T N_A 100}\ \ _{ion}}\right)$. From the polarity of the G-FET response which indicates a negative doping of the graphene sheet, we could assume the absorption or physisorption of the positively charged potassium ions on the graphene surface. This positive charge accumulation acts as a positive gate leading to an increase of the electron density and higher drain currents. These results further confirm the ability of G-FET to detect extracellular transient ion fluxes at the microscale and to provide a dynamic monitoring of the electrolyte concentration. We have then assessed in which extent G-FETs can monitor biologically gated ion fluxes by coupling G-FET array with Xenopus oocyte expressing ion channels, and investigated the potential of such biological and electronic hybrid-device for chemicals sensing.

**Coupling G-FET with non-vertebrate and artificial ICCRs.** In non-vertebrate animals, insects possess a particular family of seven transmembrane ICs (7TMICs) that mainly compose their olfactory and gustatory systems.[33] Among these 7TMICs, the insect olfactory receptors (iORs) form the largest subfamily and are composed of two distinct subunits, a highly conserved odorant receptor co-receptor (Orco) that physically and functionally associates with odorant receptor (OR) subunits that evolved to recognize specific odorant molecules. Orco is able to form homotetrameric cation channels[34] and is activated by few synthetic ligands such as the original one, VUAA1.[35] Orco generates high current amplitudes in response to ligand (VUAA1) in Xenopus oocytes,[3] and it was consequently used as a first model (figure 4 and 5). On the other hand, artificial ICs can be engineered by fusing ICs with G protein-coupled receptors (GPCRs).[4,5] GPCRs are not ICs but membrane receptors belonging to the largest family of human membrane receptors with more than 800 members and they are activated by a very large diversity of ligands from ions to proteins (e.g. hormones), as well as photons.[36] Thus the combination of ICCRs with G-FET have been later tested (figure 6), as it could further provide suitable candidates for real-time detection and identification of various biomarkers or chemical compounds in general.

For each IC models (Orco and ICCR), heterologous expression of the ICs in defolliculated Xenopus oocytes were obtained by standard micro-injection of mRNA and $\geq$ 72h incubation at 19°C in modified Barth's solution supplemented with penicillin, streptomycin and gentamycin antibiotics. Xenopus oocytes have been placed above the FET arrays that are shown on figures 4 and 6. The orientation of the oocyte over the FET is of particular importance due to the heterogenous density of ion channels gradually concentrated around the injection point of mRNA[37]. To easily identify the injection point after incubation, all oocytes were micro-injected in the animal dark pole. The vitelline envelope surrounding the oocytes and made of glycoproteins was previously peeled off to reach the lipidic plasma membrane containing the ICs. This step was performed just before the experiment due to the high fragility of peeled-oocytes, and contact with the air interface had to be strictly avoided to prevent the destruction of the membrane. The dark pole was carefully placed over the FETs devoted to recordings. The G-FET array and the oocytes were immersed in 300µL of a physiological buffer (ND96) containing high concentrations of sodium (91 mM), chloride (99 mM) and calcium (1.8 mM), and low concentration of potassium (2mM) preserving chemical gradients with intracellular concentrations (~10 mM $Na^+$, ~44mM $Cl^-$, ~$10^{-5}$ mM $Ca^{2+}$ and ~109 mM $K^+$)[38] as described within figure 1b. A large PDMS chamber assembled with the G-FET array allowed to contain the liquid and the quasi-reference Pt electrode used for applying the liquid gate voltage.

For all assays, the G-FET signal was recorded few minutes after oocyte immersion to ensure the membrane adhesion to G-FET. The drain current was monitored in time, while injecting small amounts of chemical compound that are expected to activate or inhibit the opening of the engineered K-channels. The recordings of Orco are shown in figures 4 and 5 and in figure 6 for the engineered ICCRs.

**Real-time detection of Orco IC with G-FET.** The time trace of the G-FET drain current has been monitored in real time during the activation of the Orco, and compared with two control conditions namely with oocyte but non-expressing Orco (non-injected oocyte), and without any oocyte (Figure 4). The opening of Orco ICs occurs when adding VUAA1 ligand (typ. 0.1mM) to the solution, which bind directly on the ICs. Figure 4a shows the drain current time trace after three injections of 30µL of 1mM VUAA1 stock sequentially increasing the final concentration of VUAA1 from 90µM to 173 µM to 250µM. These three successive injections result in clear increases of the drain current $\Delta I$~23 nA, 21 nA and 10 nA with a signal-to-noise ratio of about S/N~9, 8 and 4 respectively. These significant increases are not visible on negative controls with oocytes that do not express Orco, neither on those without oocyte coupling confirming further the detection of Orco channel activation and the ability of G-FETs to measure transmembrane ion fluxes from few ion-channels localized at the cell-device interface.

The opening of Orco ICs should induce an outward flow of $K^+$ and an inward flow of $Na^+$ and in lesser extent $Ca^{2+}$ (Figure 4d). The inward flow of micromolar concentration of $Ca^{2+}$ could then activate endogenous calcium-activated chloride channels $Cl^-_{Ca}$ (inward $Cl^-$ fluxes) which is usually not observed in TEVC recording because the equilibrium potential of $Cl^-$ ($E_{Cl^-}$ = -20mV) is slightly higher or close to the measure $V_m$. Thus, the flow of $Cl^-$ is usually negligible when biasing the membrane potential around -20mV (inset Figure 4a). However, such current could occur during G-FET measurement as the membrane potential is floating (Figure 4a) or when applying higher negative value of the membrane potential (Figure 5c).

The graphene channel is slightly n-doped when interfacing with the oocyte (Figure S3). In the electron regime, the polarity of the G-FET response ($\Delta I_D$ > 0, Figure 4a) is as expected for sensing positively charged cations, which are presumably potassium here. Indeed, the concentration of $Na^+$ being significantly higher than $K^+$ in the extracellular medium (91 mM NaCl and 2 mM KCl, see material and methods), an increase of $K^+$ is expected to be easier to be detected. Also, the G-FET signal is completely suppressed when blocking $K^+$ current at $V_m=-80mV$ (Figure 5c). Interestingly a double peak occurs for each injection which could stem from the activation of another type of ICs, possibly inward calcium $Ca^{2+}$ that is present at low concentration in the extracellular bath (1.8 mM $CaCl_2$).

Despite the double peak, the time response is similar to what was obtained by inducing artificial ionic fluxes (see figure 3a). The drain current increases first toward maximal values and then decreases toward a lower drain-current value after the complete diffusion and screening of the additional charges induced by the activations of the ICs. Note that the current step between VUAA1 applications slightly decreases while adding new agonist. Indeed, a large number of Orco channels are open after the first agonist application. Following VUAA1 injections will increase slightly the number of open ICs until it reaches a saturation state.

**Dual TEVC and G-FET recordings of ligand-gated Orco.** To further assess the field-effect transduction of physiological ion fluxes, real-time TEVC and G-FET were performed simultaneously (figure 5). The Xenopus oocytes were placed on G-FET array immersed in PDMS chamber containing 300µL of ND96 and were impaled by two pulled pipettes (~0.22 MΩ) filled with 3M KCl and an Ag/AgCl electrode measured against a reference electrode immersed in the ND96 buffer surrounding the oocytes (same condition than previously). Initially, the oocytes exhibited a resting membrane potential around -15mV and the membrane voltage was clamped at -10mv (Figure 5a) and -80 mV (Figure 5c). On figure 5a, the simultaneous time traces of the TEVC and G-FET recordings reveal a small increase of drain current while adding 30µL at 1mM of VUAA1 (90µM final at equilibrium state). The signal amplitude is comparable to previous measurements (figure 4). However, the signal-to-noise ratio is lower and probably due to noise injection of TEVC coupling. The current measured by TEVC is also increasing toward negative values, which confirms the activation of Orco but indicates a dominant inward flow of positive charges across the cell membrane mainly induced by $Na^+$ ions, which are counterbalanced by an outward flow of K+ ions (Figure 5d). Without the outward $K^+$ current, the TEVC signal is expected to be higher (few µA, inset figure 4a). This result indicates that G-FET dominantly detects the increase of $K^+$ ions in the extracellular space (which contains only 2mM of $K^+$, $\Delta C_K = +107\ mM$) that is proportionally higher than the depletion of $Na^+$ from the buffer (which already contains 96mM of $Na^+$). It should be notice that the current variations are much more significant (20 times higher) in the TEVC than in the G-FET recording. While the current is collected from the whole cell during TEVC recording, the reduction of the sensing area when using G-FET (from 1mm² to 20µm², respectively) should reduce the number of detectable ICs and the amount of potassium ions detected by G-FET. The variation of drain-current is thus highly dependent with the IC heterologous expression rate. Considering a typical density of $10^5$ IC.mm$^{-2}$, the number of ICs in vicinity of the graphene surface should be about 10 ICs (versus ~$10^5$ for whole cell TEVC recordings). A slight variation of the ion channel density could significantly impact the detection efficiency of the G-FET measurement. Although TEVC exhibits clear response to ligand activation, the G-FET detection of ion currents is not reachable on every measured oocyte, depending mainly on the expression rate of ICs, the density of ICs coupled with the graphene channel and the quality of the interface after successive uses.

When clamping the voltage at -80mV (figure 5c), the outward K+ currents are blocked. As a result, the amplitude of the TEVC current is increased (330nA at -10mV to 1500 nA at -80mV), which is expected with higher inward flow of Na$^+$ and Ca$^{2+}$ and a lower outward flow of K$^+$ (red trace figure 5c, blue arrow). The absence of response when adding the ligand (black curve figure 5c) confirms that the G-FET is rather sensitive to the outward potassium current than to the sodium entry. Interestingly, after few minutes the G-FET response reveals drain current variations with opposite polarity (figure 5c, red arrows). This inverse polarity agrees with a rapid increase of Cl$^-$ ions in the extracellular space when the membrane voltage is clamped at -80mV (Figure 5e). Xenopus oocytes expressed Ca$^{2+}$-activated chloride (Cl$_{Ca}$) channels that generate large Cl$^-$ currents when the concentration of intracellular calcium ions increases in micromolar range (EC$_{50}$ ~25 µM)[39]. When Orco channels are activated, a large flow of Ca$^{2+}$ enters into the cell due to the high gradient and the high negative membrane voltage, and Ca$^{2+}$ ions activate these Cl$_{Ca}$ channels. Due to the important density of these channels especially in the dark animal pole in contact with graphene[39] and their sensitivity to intracellular calcium and the weak outward flow of K$^+$ at V$_m$ = -80 mV, Cl$^-$ accumulation in graphene vicinity is dominant. The nature of the signals is also consistent with the inversion of polarity on G-FET measurements. A Cl$^-$ accumulation at the graphene-cell interface induces a positive p-doping of the graphene channel, reducing the G-FET channel conductance in the electron regime. Here, G-FET exhibits clear response to Cl$^-$ flow. The recordings enable to identify the ion polarity of transmembrane currents (positive out or negative in) thanks to signal polarity, which is not the case for TEVC. Indeed, negative variations of TEVC current are induced either by positive charges flowing inward the cell or negative charges going outward.

**Application for the G-FET detection of ICCRs.** The designed ion channel coupled receptor is activated by muscarinic M2 receptor agonists such as acetylcholine (ACh) and antagonized by compounds such as atropine (At). The signal is generated by opening of the coupled potassium channel and it is blocked by barium (B). The experiments were performed on the G-FET setup (figure 6a) that was not parallelized with the TEVC setup. The drain current was monitored in time, while injecting 3 µL of ACh, At and B, that are expected to activate and inhibit the opening of the coupled $K^+$ channels, and then block all cell channels respectively. Figure 6 shows successive timestamps of about 1 to 5 minutes, which correspond to acquisitions performed at a sampling rate of 4 kHz.

Prior any injection, the baseline value of the drain current is about 53 µA ($t = 50$ s figure 6a, with $V_{DS} = 90$ mV, at $V_{LG} = 0V$) which is as expected from the $I_D$-$V_{LG}$ curve performed just before placing the oocyte (Figure S3). The addition of acetylcholine (ACh, 5 µM final) induces a sharp increase of the drain current $\Delta I/I_0 \sim 0.6$ and a constant rate of about 0.8 µA·s$^{-1}$ (figure 6b). This endogenous ligand binds and activates the human M2 muscarinic receptor, which opens the coupled potassium channel (Kir6.2) and generates transmembrane potassium outward flow. This resulting $K^+$ flux toward the extracellular medium modulates the underlying G-FET conductance. The polarity of the G-FET response ($\Delta I_D > 0$) is the same as previous detection with Orco ICs. This observation validates both the expression of the receptor coupled ion channel within the oocytes, and the efficient detection of the ICCRs with G-FET.

The subsequent addition of atropine - a potent M2 receptor antagonist - immediately blocked almost totally the activation of the receptor that returned to a state close to its basal state and induced the same change of state of the coupled-$K^+$ channel. The drain current remained relatively constant few seconds before fluctuations in lower values. These variations, ranging from 500 nA to 1-5 µA, could result from the activation of endogenous $Na^+$/$K^+$ ATPases that extracellularly export $Na^+$ and intracellularly import $K^+$ ions and certainly participate in the variations of the density of positive charges in the inter-membrane/G-FET space in coordination with IC activity. The last injection of barium (3mM final) blocked the potassium channels and suppressed all fluctuations. The baseline of drain current remained constant up to the end of the experiment ($I_D = 89µA$). Note that the time trace still exhibited step-like variations $\Delta I_D$ ranging between 70 nA and 130 nA (figure 6d). Theses fluctuations are not observed prior the activation of the K-channels by ACh (figure 6e) and could stem from the flickering noise of few ICCRs, since the rectangular type signals are similar in single ion channel recordings.

From the amplitude of the drain current, we could estimate the among of ICs sensed by the G-FET. The following equation enables to extract the equivalent variation of gate voltage within the cleft between the oocyte and the graphene channel $\Delta V_g = \Delta I_D \times (g_m \cdot V_D)^{-1}$. Considering the normalized transconductance $g_m$ (1mS/V) and the drain-source voltage $V_{DS}$ (90mV), the equivalent variation of the gating potential $\Delta V_g$ in the cleft ranges between 5 and 15 mV (for drain current steps of 0.5 and 1.5 μA) after injecting the atropine. For low surface potential, the Grahame equation provides the equivalent number of surface charges (or ions) by $\Delta n = (A \varepsilon_e \varepsilon_0 / \lambda_D) \times \Delta V_g$. This value is ranging here between Δn=0.6 pC and 1.8 pC, with the exposed graphene area and Debye length being A=200μm² and $\lambda_D = 1\ nm$. Single ICCR current being about 1-3 pC·s⁻¹, the amplitudes of the step-like variation are indeed consistent with the opening/closing of individual ICCRs.

This results with ICCRs further demonstrated the high sensitivity and versatility of the G-FET detection for sensing transmembrane current in biological systems, which is a key building blocks for implementing advanced bio-hybrid electronics.

**DISCUSSION**

Here, the reported sensitivity and limit of detection have been shown to be comparable to state-of-the-art functionalized graphene devices, but with an additional advantage which is the ability to measure several ion species (e.g. $Na^+$, $K^+$, $Cl^-$) in real-time with a same G-FETs. This feature is crucial for electrophysiological recording and for advanced biohybrid electronics. Indeed, this first demonstration of combining graphene ISFET with biological ICCRs open avenue of investigation, regarding the available number of human ICs that might be provided by the ICCRs (up to 800 members).

Graphene bioelectronics has demonstrated efficient sensing properties enabling to detect few to single ligand-activated ICs distributed on very localized surface of the oocyte membranes. However, a current drawback of the field-effect detection with graphene remains the significant number of undetected events (ligand-activation of ICs) which could be an issue for high throughput biosensing. To overcome this possible limitation, several features might be optimized to provide the required detection rate and reproducibility. Here, we discuss several device parameters that could be improved in future development, such as the detection threshold, the sensitivity, the signal-to-noise ratio, and the quality of the interface.

Because graphene surface charges are rapidly screened by charges in bulk solution, the limit of detection of the reported G-FETs is quite high when sensing ionic species in large volumes (~$2.10^{-8}$mol in 200μL, figure 3). However, strong coupling between the G-FET and the ion-channel receptor, such as in Xenopus oocytes, enables to sense weak variations of ion concentration in time (being around $10^{-16}$mol per second for 20 opened ICs). Thus, the sealing resistance at the cleft between the oocyte and the graphene channel is expected to improve the detection sensitivity. In this closed configuration, small variations of ion amount are easier to detect than in an opened media. Although this apparent high detection sensitivity, the response of the G-FETs could be further improved by tuning the G-FET operation regime toward more sensitive setpoints (e.g. $V_{LG}$= -500mV, figure 3a) without altering the physiological properties of the oocytes and the ICs to-be-sensed.

Beyond the amplitude of the G-FET response, the electrical noise is an important feature to improve the number of detected events. For G-FET recordings, the typical noise amplitude is around 3 nA, enabling good signal-to-noise ratio (S/N~9) while detecting ionic currents (figure 4). However, the noise amplitude was significantly increased when coupling G-FET to TEVC (figure 5), reducing further the detection efficiency. Dedicated setup that will reduce the background noise during dual G-FET and TEVC recording, are thus expected to further enhance the detection rate.

On the other hand, the G-FET detection efficiency is expected to be lower than whole cell TEVC-measurements as it mainly depends on the probability of coupling G-FET with a well activated IC. Indeed, the ligand application might trigger the opening of ICs which are located away from the graphene channel and not detected by the G-FET. To overcome this limitation, experimental conditions can be improved on both biological and electrical aspects. To that end, expression rate of ICs has been optimized ($100.10^3$ IC.mm$^{-2}$). Moreover, the oocytes are interfaced with G-FET by coupling the pole with higher expression probability (dark pole). Despite this precaution, the number of coupled ICs is expected to be significantly reduced, being about 10 for a 10×20 μm² device if all ICs are activated and if the ICs density is maximal and homogeneous above the G-FET. For such low quantity, the stochasticity of the ICs expression or activation could no more be negligible. Thus, the number of activated ICs above the GFETs is probably less than 10 ICs. While reducing the device size might be a strategy for sensing individual or few ion channels, some applications may require higher efficiency. In that case, the ICs are also more likely to be sensed by increasing the effective surface of graphene-membrane coupling. Larger graphene devices can be designed to increase the detection surface and the probability to interface ligand-activated ICs with the graphene channel. Then, the G-FET dimensions should be carefully chosen to keep the required sensitivity which is proportional to W/L ratio, W and L being respectively the width and length of the transistor.

Here, five independent G-FETs were used to provide first demonstrations within liquids and cells. Future investigation should be done to optimize the demonstration in show case application, providing statistical study of the sensing performance and assessing the robustness of the calibration which are both crucial for chemical detection.

Another asset of arrays of micrometer-size G-FET is multisite and multi-cell recording. Hence several areas of the oocytes or several oocytes could be measured at the same time, which optimizes the detection probability. Additionally, G-FET arrays can be coupled to microfluidic circuits to control the oocyte/G-FETs coupling and liquid injections for drugs applications or medium changes.

Lastly, the chemistry at the graphene-liquid interface is suspected to play a key role in the detection efficiency and repeatability. The hydrophobic monolayer graphene strongly interacts with the surrounding proteins, as well as with the cell membrane. The absorption of phospholipids at the graphene surface could act as a physical barrier for the ions to be detected. Also, the presence of residual chemicals such as IC-inhibitors (e.g. atropine, baryum) at the graphene surface will block consecutive activations of ICs, thus no signal will be detected when repeating the experiment with the same G-FET. Thus, particular care will be taken in future works to assess the extent of such interaction on the detection rate. Dedicated rinsing protocol and functionalization strategy could then be implemented to optimize the interaction with the ions to-be-detected while minimizing the reaction with the other surrounding biological or chemical species.

**CONCLUSION**

The reported works demonstrated the high sensitivity of the G-FET for the detection of ultralow variation of ion concentration in saline media. The reachable detection threshold being about few to single potassium $K^+$ ICs, G-FETs outperform conventional microdevices while keeping an optical access to the recording site for interfacing soft and living matter. To that aim, the graphene field effect transistors have been successfully assembled with engineered Xenopus oocytes for the detection of ligand-activated ICs. The G-FETs have been shown to efficiently sense the evoked activity of ICs by ligand application in-operando (cell culture solution) and for two classes of ICs (Orco and ICCR). Dual-acquisitions with TEVC and G-FET demonstrated that a G-FET enables efficient transduction of ion fluxes into readable electronic signal, as clear similitudes have been observed in the time traces of the TEVC and G-FET signals. The electronic properties of liquid gated G-FETs allowed quantitative measurements in high frequency regime that resolved a broad dynamic of ICs opening and closing. Beyond the detection of transmembrane current ions, G-FET measurement could allow to identify the polarity of the sensed ions, providing additional cue on the nature of the ion fluxes to complement TEVC measurements. Finally, the micrometer size of G-FET enabled local measurement of few individual ICs, in addition of being sensitive to whole cell current variation. The multisite recording ability offered by G-FET array rises also numerous possibilities for multiscale sensing, which is of both fundamental and applied interests such to provide functional map over single or assembly of cells, or for high throughput screening of cellular solutions or analytes. Although the promising sensing properties of G-FET have been reported, the measurement repeatability should be further improved to bring this novel kind of biohybrid devices as the next golden standard for ICs recordings.

**Figure 1. Principle of hybrid bioelectronic sensing.** (a) The schematics describe a graphene field effect transistor interfacing a Xenopus oocyte heterologously expressing engineered ion channels. The chemicals to be-sensed (teal disk, labelled ligand) bind to the receptor part (red) and trigger the opening of transmembrane ion channel (purple), which results in a change of the ion flow through the channels. Following the ligand-activation of ICs, the charges in excess ($K^+$ mainly, blue + circles) above the graphene transistors channel can be detected by monitoring the G-FET drain current. On the right, the expected shift of the G-FET transfer curve ($I_D$-$V_{LG}$ curve) for both accumulation (+Q) or depletion (-Q) of positive ions. (b) Schematic zooming on the interface between the oocyte membrane and the graphene sheet. Model of ICs are shown as ICCRs that are depicted in ovoid-shape with a receptor moiety in red and recognizing ligands in green, and with an ion channel moiety in blue allowing ions to pass through the membrane (Mb). Endogenous calcium-activated chloride channels ($Cl_{Ca}$) are depicted in purple. Concentrations of cations are indicated in the intracellular space (Int) or extracellular (Ext) in the ND96 buffer and relative theoretical equilibrium potentials are indicated on the right.

**Figure 2. Manufacturing graphene field-effect transistors.** (a) Optical micrograph of a 20x5 $\mu m^2$ G-FET after fabrication process. An arrow indicates the position of the graphene transistors channel which is directly exposed to the analyte or bio-object to be sensed. The drain and source metallic leads (brightness areas) are encapsulated with an insulating resist for liquid operations. (b) Atomic force micrograph of a G-FET channel after passivation. A cross section exhibits the characteristic heigth of wrinkles and impurities (1-5nm). (c) Raman spectroscopy averaged (red curve) on 5 spatial points (grey curves) within a G-FET channel. (d) Transfer characteristic and transconductance of a G-FET measured in PBS.

**Figure 3. Ion detection performance of the graphene field-effect transistors.** (a) $I_D$-$V_{LG}$ transfer curves of a 20×10 μm² G-FET as function of the ionic concentration of the liquid gate, at fixed drain-source voltage ($V_{DS}$ = 80 mV). The liquid gate voltage $V_{LG}$ is applied with a quasi-Pt reference electrode immersed in the solution. The concentration of KCl is increasing from 0 to 300 mM (yellow to blue respectively), leading to shift the position of the Dirac point toward negative values. (b), (c) Evolution of the drain current $I_D$ over time while adding small amounts of KCl stock. The added KCl amount is indicated in blue. The electrolyte potential is fixed at (b) $V_{LG}$= 400mV and (c) -500mV. (d) Dirac shift $\Delta V_D = V_D - V_{D|C=0}$ corresponds to the value of the Dirac point compared to the value obtained at [Cion] = 0 M as a function of the KCl concentration, for two different G-FETs (blue and red). The slope corresponds to the G-FET sensitivity to variations of ionic concentration. (e) Drain current variations $\Delta I_{DS} = I_{DS} - I_{DS|C=0}$ compared to the initial drain current at [$C_{ion}$] = 0 M as function of the KCl concentration. Blue points are extracted from the drain current value at fixed $V_{LG}$ = 400 mV on the transfer curves (from figure 3a). Red points correspond to the on real-time stabilized drain current value after KCl injection (from figure 3b).

**Figure 4. Real-time detection of Orco channel activation in Xenopus oocyte.** (a) Timestamp of the drain current $I_D$ during the three successive applications (30μL) of Orco agonist VUAA1 (respectively 90μM, 183μM and 260μM). Same ligand application is repeated on non-injected oocyte (does not express the Orco protein) and G-FET without oocyte coupling. G-FET is polarized at $V_{DS}$ = -20mV. Acquisition frequency is 25kHz. Inset: Typical response of two-electrode voltage clamp (TEVC) to VUAA1 activation of Orco channels. By convention, the current is negative, and an increase of the current amplitude results in a downward deflection. (b) G-FET electronic set-up under transmission microscopy. The reference potential of buffer electrolyte (ND96) is fixed with a quasi-Pt reference electrode at Vg=0V. (c) The Xenopus oocyte is placed on a G-FET array using optical microscopy. G-FET used for Orco real-time measurements (a) is marked by a red circle. (d) Schematic showing the flow of ions through Orco expressed in the plasma membrane of oocytes and when activated by VUAA1. Endogenous $Cl_{Ca}$ channels are depicted in purple and activated by intracellular calcium. $V_m$ is the transmembrane voltage of the Xenopus oocyte at the resting state.

**Figure 5. Real-time measurements of G-FET and TEVC signal while activating Orco channels in Xenopus oocyte.** (a) Representative coupled TEVC – G-FET recordings showing the currents induced by Orco agonist (VUAA1). The membrane potential is $V_m$ = -10mV. G-

FET is polarized at $V_{DS}$ = 10mV. Both G-FET (blue) and TEVC (red) responses to VUAA1 application are characterised by an increase of measured current. Dash line corresponds to the TEVC current baseline (b) Coupled G-FET and TEVC measurement set-up. The membrane potential of the oocyte is monitored by TEVC. (c) Real-time recording of ionic currents with higher amplitude. TEVC current induced by VUUA1 application is indicated by the blue arrow. Red arrows indicate transient ion currents. $V_m$ = -80mV, $V_{DS}$ = 20mV. (d) Schematic showing the flow of ions through Orco and $Cl_{Ca}$ when VUAA1 is applied and the transmembrane voltage clamped at -10mV. (e) Same schematic as d) but with a transmembrane voltage clamped at -80mV.

**Figure 6. Real-time detection of engineered K+ channels expressed in Xenopus oocyte with G-FET.** (a) Optical micrograph of the sensed oocyte which is directly positioned on a G-FET. (b) Successive timestamps of the drain current $I_D$ during the successive application of acetylcholine (ACh), the endogenous ligand, then atropine (At, antagonist) and finally barium (B, K[+] channel blocker) as highlighted by the three arrows. (c-d) Zoom-views of the $I_D$ time course, just after adding atropine (c) which is expected to inhibit the ACh activation of the ICCRs and after the injection of barium (d) which suppresses all signals. (e) Control current time trace recorded prior any injection on same G-FET coupled with the oocyte. There is no variation of ion concentration at the G-FET interface. (f) Schematic showing the flow of potassium ions through the muscarinic M2 ICCR when activated by acetylcholine (ACh). Atropine (At) is a potent antagonist that inhibits the activation of M2. $V_m$ is the transmembrane voltage of the Xenopus oocyte at the resting state. Barium (B) is generic blocker of potassium ions.

**METHODS**

*G-FET fabrication and characterization.* Monolayer graphene was obtained by chemical vapor deposition (CVD) and transferred from the copper growth substrate onto sapphire by wet transfer technique. The frontside of the copper foil supporting the monolayer was protected by a thin layer of PMMA resist, while the backside was successively etched in oxygen plasma (few seconds) and ammonium persulfate (1mM, for about 1h). After complete copper etching, the floating graphene/PMMA bilayers was rinsed in several baths of DI water to remove all solvent traces and then gently fished with the final substrate (to-be-used for the sensing experiments). The sample was left 1 hour under a laminar flow hood to dry, and finally heated up to 180°C to remove water and improve the adhesion of the monolayer graphene on the substrate. The graphene channels were then obtained by patterning a positive resist with laser lithography and a dry etching in $O_2$ plasma. The metallic drain and source contacts were obtained by wet etching of metals (1 nm Ti, 30 nm Pd, 20 nm Au) design using a negative resist mask. These metallic lines were finally passivated with S1818 resist to insulate them from the liquid gate (Figure 2a). Figure 2 shows typical sample and zooms on a G-FET to illustrate the high quality of the resulting graphene channels directly exposed to the analytes. Only few residuals of the fabrication process are observed, and wrinkles can already be observed as highlighted by the arrows.

*Expression of the ICs in Xenopus oocytes.* Orco gene from *Drosophila melanogaster* was codon-optimized, synthetized and subcloned in derived pGEMHE vector by Genscript. pGEMHE was optimized for protein expression in Xenopus oocytes as previously described.[6,31] To design ICCR, the genes of human M2 and mouse Kir6.2 was fused by PCR by insertion of the GPCR coding sequence at the 5' end of the Kir6.2 gene cloned in derived-pGEMHE vector. DNAs coding for Orco and ICCR were amplified using Qiagen MidiPrep Kit, linearized in the 3' region of the polyA tail, purified by standard phenol:chloroform extraction and transcribed in mRNA using the T7 mMessage mMachine Kit (Thermo Fisher Scientific) following the supplier's protocol. mRNAs were purified by the standard phenol:chloroform protocol, analyzed by agarose-gel electrophoresis and quantified by spectrophotometry[34]. For the heterologous expression[26], *Xenopus* oocytes were defoliculated by collagenase treatment and microinjected with 50 nl of water containing 20 ng of Orco or 5 ng of M2-Kir6.2 cRNAs. Oocytes were incubated in individual wells in Barth's solution supplemented with penicillin and streptomycin and gentamicin mix for at least 48 hrs at 19°C. Recording buffer (ND96) was composed of 91 mM NaCl, 2 mM KCl, 1.8 mM CaCl2, 1 mM MgCl2, 5 mM HEPES, pH 7.4. In ICCR recording, 0.3 mM niflumic acid was added in the buffer to partially block endogenous Cl⁻ currents. Before loading oocytes on G-FET, the vitelline envelope of Xenopus oocytes was manually removed with sharp forceps (Moria 9980) under a stereomicroscope (Leica MZ6) and in ND96 buffer.

All chemicals were purchased from Sigma-Aldrich. Stock solutions were acetylcholine chloride 5mM (water), atropine 1mM (ethanol), $BaCl_2$ 1.5M (water) and VUAA1 108mM (DMSO).

*TEVC and G-FET sensing recordings.* For parallelization of G-FET and TEVC recordings, the G-FET array and the electronic card were placed in a standard manual TEVC setup in a Faraday cage and under a stereomicroscope (Leica Wild M3Z). In a PDMS chamber containing 300µL of ND96 buffer, the peeled oocytes were placed with a Pasteur pipette on a connected G-FET that was electrically characterized before oocyte addition. Using micromanipulators, the oocytes were impaled by two pulled borosilicate pipettes (WPI TW150F-4, ~0.22 MΩ) filled with 3M KCl and an Ag/AgCl electrode. Reference electrodes are immersed in the ND96 buffer surrounding the oocytes. The TEVC setup was composed of the Geneclamp 500B amplifier (Axon), the Digidata 1440A (Axon) and controlled with Clampex 10.44 software (Axon). Recordings were performed at 10 kHz and filtered at 10 Hz with a voltage clamped at -10mV and -80mV (figure 5a and 5b respectively). Ligands were applied successively at the appropriate volumes with a standard micropipette. The electronic card allows to record simultaneously two independent G-FETs on the same chip. Ion-sensing within KCl solution was performed on two independent G-FETs. Three independent G-FETs were used for ion-sensing within living cells.

## Supporting Information

Supporting Information is available from the Wiley Online Library or from the author.

## Acknowledgements

Authors thank Hervé Pointu, Soumalamaya Bama Toupet and Charlène Caloud for the management and the maintenance of Xenopus; Bruno Fernandez for the equipment maintenance and assistance in the NANOFAB clean room facility; Gael Moireau and Pierre Gasner for their assistance in the management of cell culture room BIOFAB where the experiments were realized. Authors acknowledge grants from the French National Agency of scientific Research under the projects ANR-18-CE42-0003 NANOMESH and from the Laboratoire d'excellence LANEF in Grenoble (ANR-10-LABX-51-01), from GRAL, the Grenoble Alliance for Integrated Structural & Cell Biology, a programme from the Chemistry Biology Health Graduate School of University Grenoble Alpes (ANR-17-EURE-0003), and from the European Research Council (ERC) under the European Union's Horizon 2020 research and innovation programme (grant agreement No 682286). IBS acknowledges integration into the Interdisciplinary Research Institute of Grenoble (IRIG, CEA).

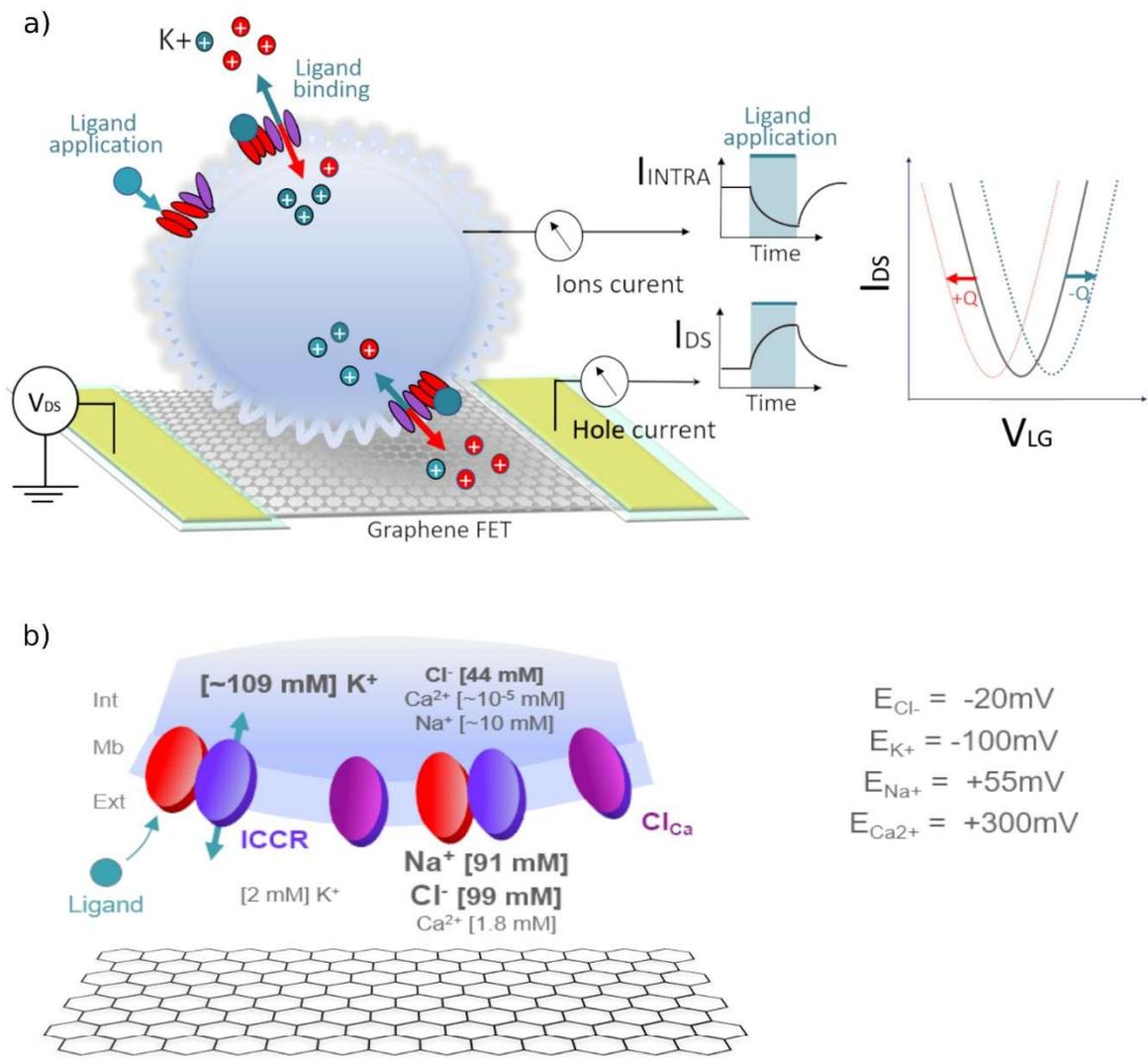

Figure 1.

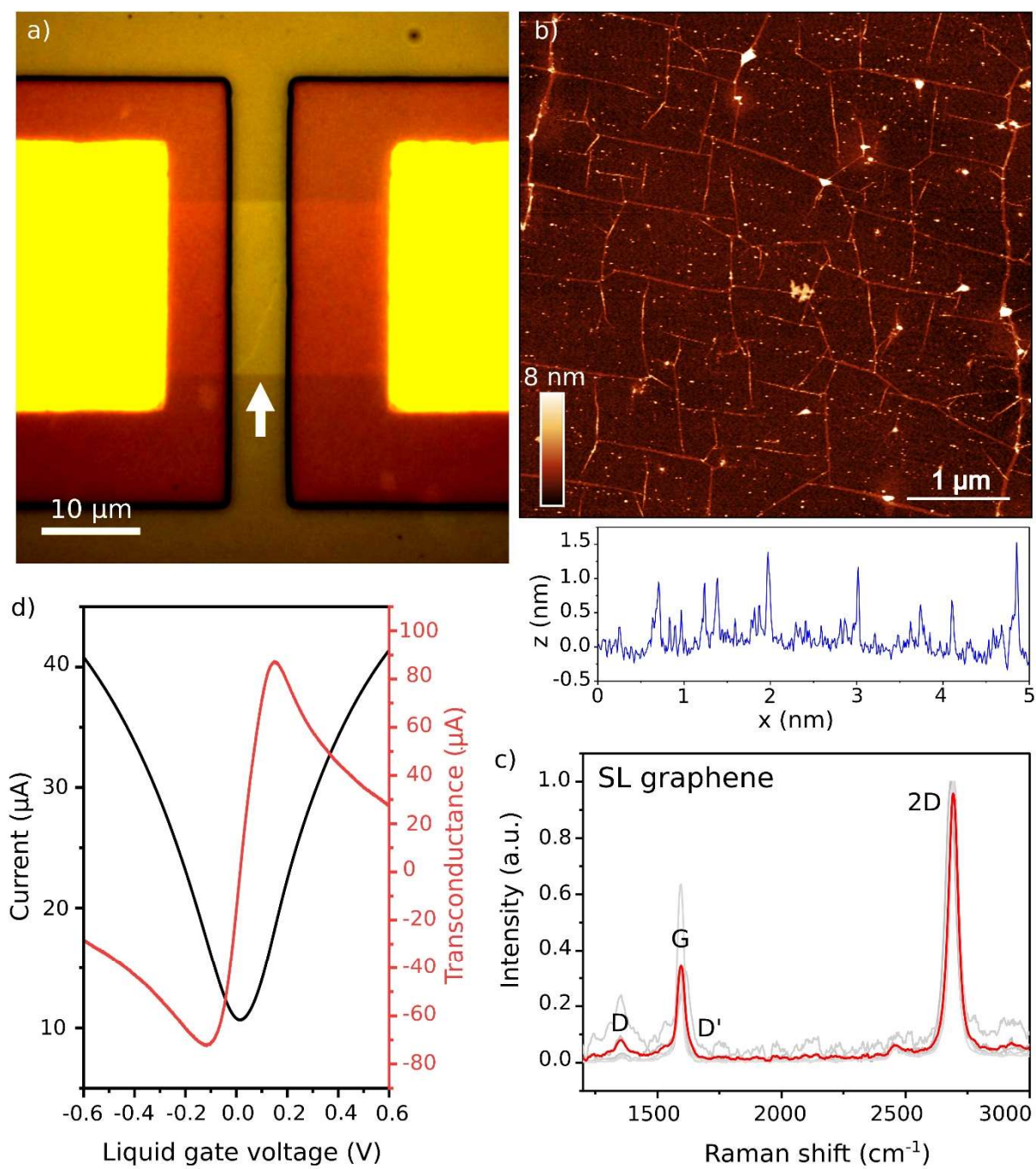

Figure 2.

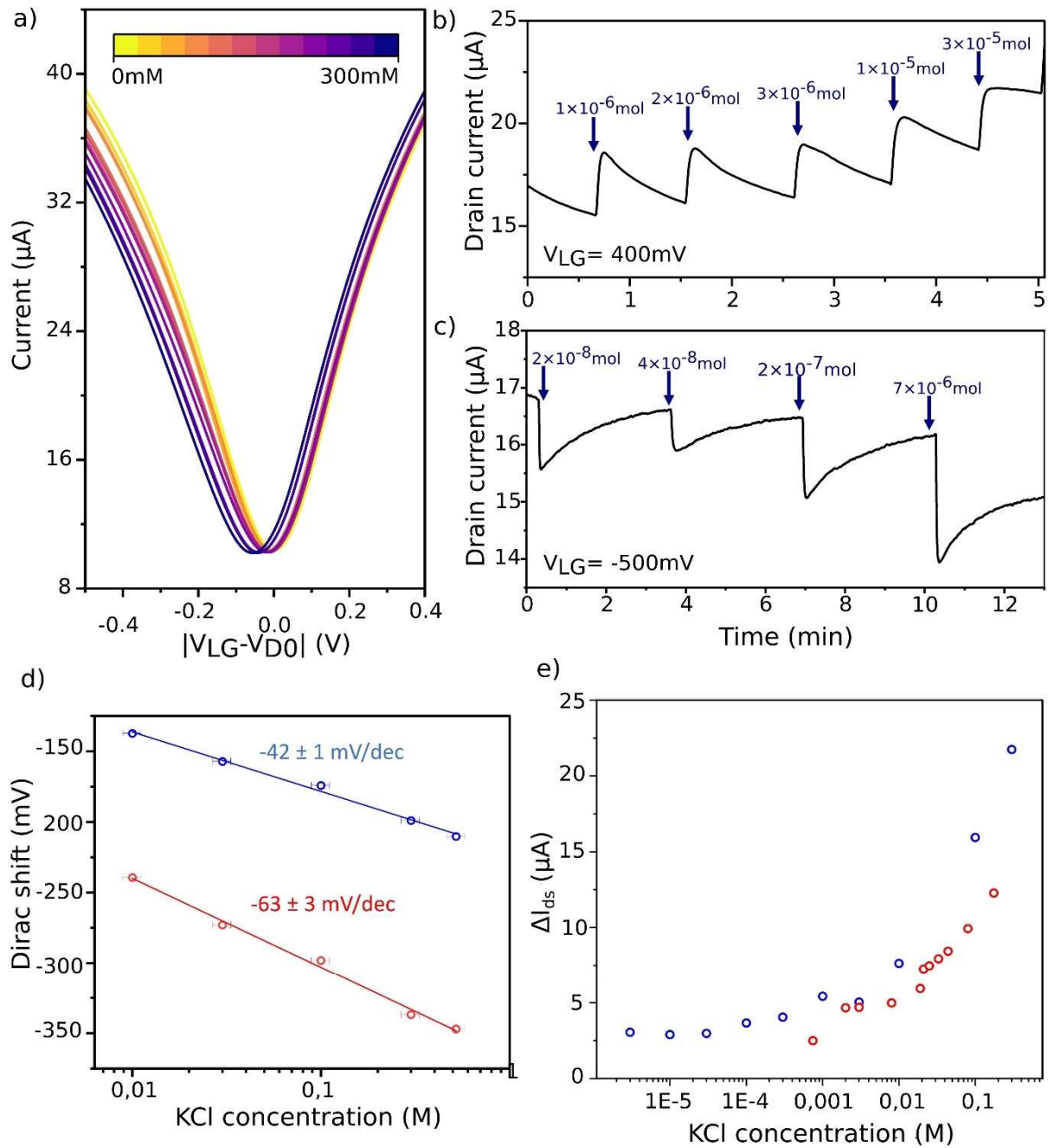

Figure 3.

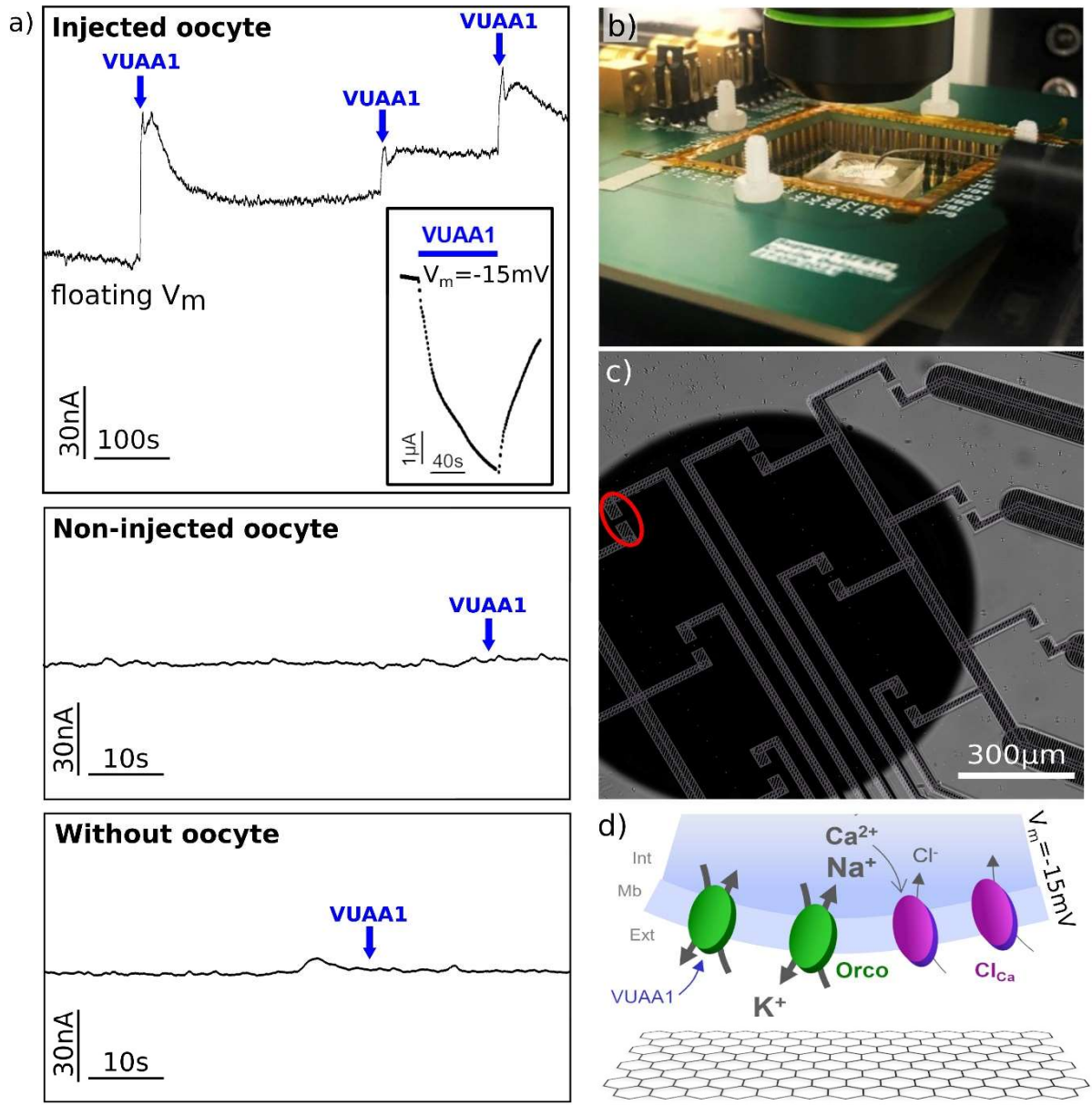

a) **Injected oocyte**

VUAA1  VUAA1  VUAA1

floating $V_m$

30nA  100s

VUAA1
$V_m=-15mV$
1µA  40s

**Non-injected oocyte**

VUAA1

30nA  10s

**Without oocyte**

VUAA1

30nA  10s

b)

c)

300µm

d) $V_m=-15mV$

Ca$^{2+}$
Na$^+$
Cl$^-$

Int
Mb
Ext

VUAA1  Orco  Cl$_{Ca}$

K$^+$

Figure 4.

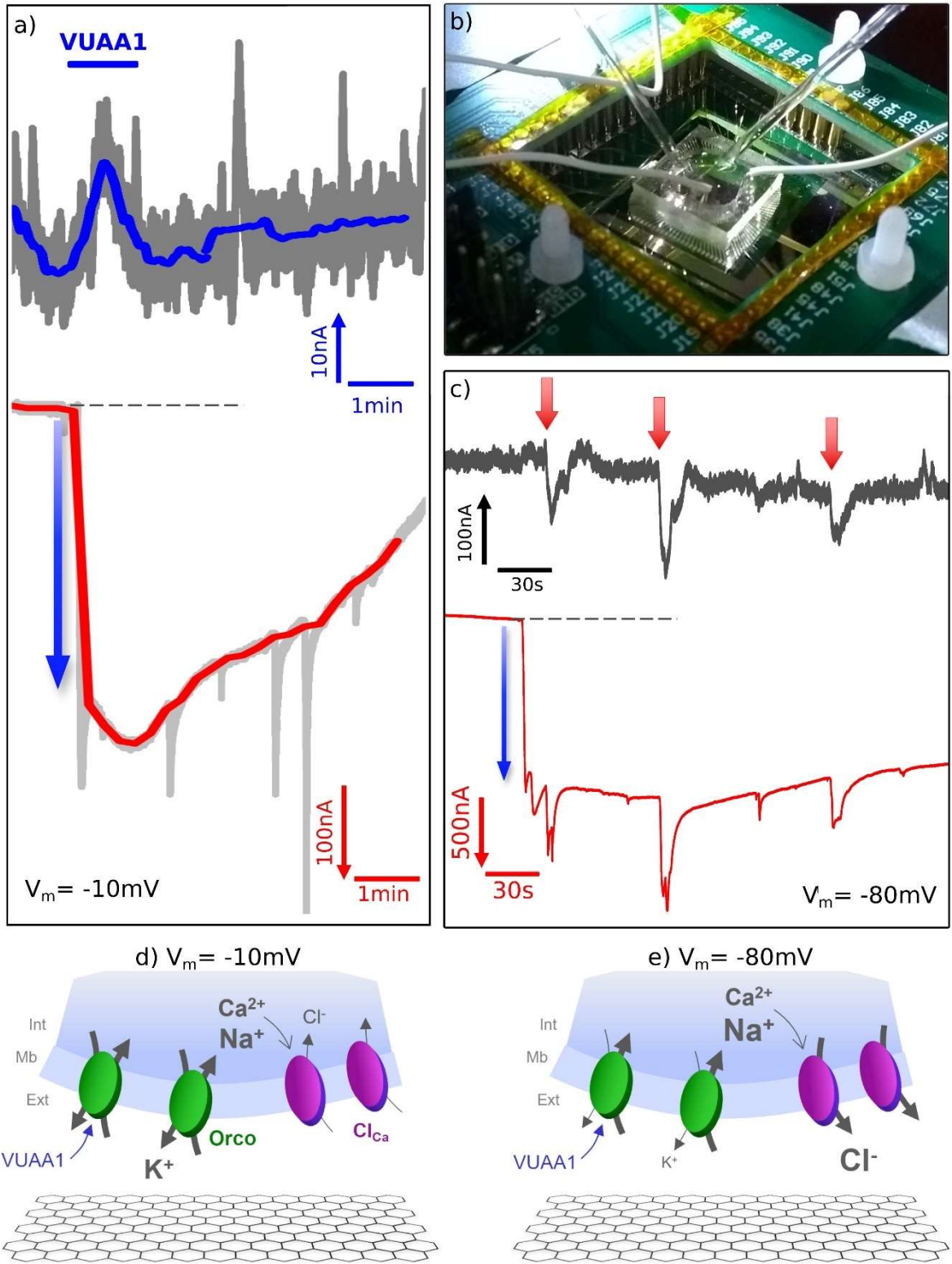

Figure 5.

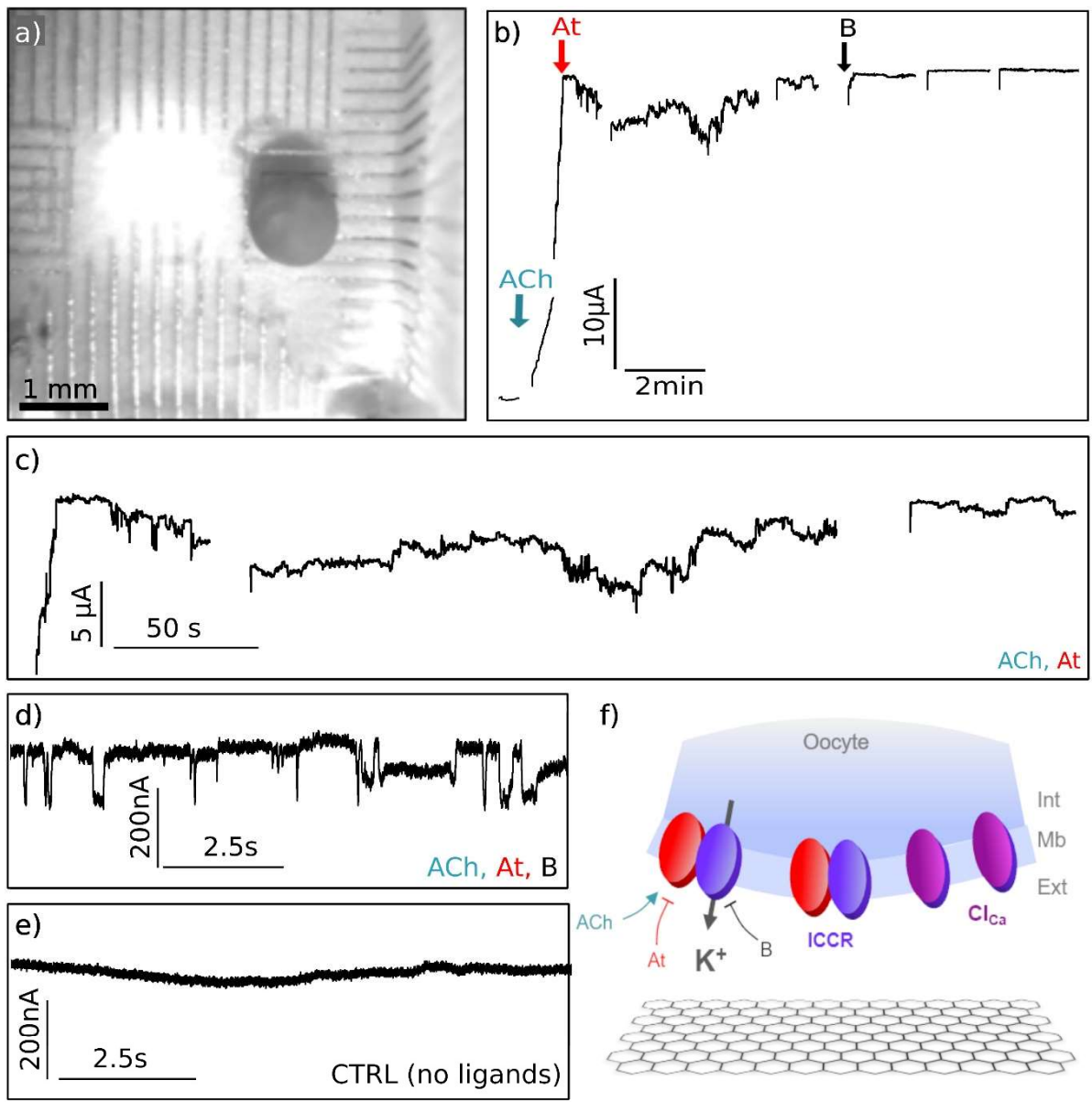

Figure 6.